\documentclass[journal]{IEEEtran}
\usepackage{cite}
\ifCLASSINFOpdf
\else
\fi
\usepackage[cmex10]{amsmath}
\usepackage{array}
\usepackage{mdwmath}
\usepackage{mdwtab}
\usepackage{graphicx}
\usepackage{euscript}
\usepackage{amsfonts}
\usepackage{bigstrut}
\usepackage{tabularx}
\usepackage{cite}
\usepackage{amsmath}
\usepackage{algorithm}
\usepackage[noend]{algpseudocode}
\makeatletter
\def\BState{\State\hskip-\ALG@thistlm}
\makeatother
\usepackage{enumitem}
\usepackage{upgreek}
\usepackage{dsfont}
\usepackage{amsmath}
\usepackage{amsthm}
\usepackage{amssymb}
\usepackage{xcolor}
\usepackage{mathtools}

\usepackage{hyperref}
\usepackage{epstopdf} 
\newcommand\numeq[1]%
{\stackrel{\scriptscriptstyle(\mkern-1.5mu#1\mkern-1.5mu)}{=}}
\usepackage{bbm}

\newcommand\LoS{\vcenter{\hbox{\scalebox{0.5}{$\mathrm{LoS}$}}}}
\newcommand\NLoS{\vcenter{\hbox{\scalebox{0.5}{$\mathrm{NLoS}$}}}}

\begin{document}

\title{Improving IoT-over-Satellite Connectivity\\ using Frame Repetition Technique}

\author{\IEEEauthorblockN {}
\IEEEauthorblockA{School of Engineering, RMIT University, Melbourne, Australia 3000
}
}
\author{\IEEEauthorblockN {Bisma Manzoor,~\IEEEmembership{Member,~IEEE}, Akram Al-Hourani,~\IEEEmembership{Senior Member,~IEEE},\\ and Bassel Al~Homssi,~\IEEEmembership{Member,~IEEE}.\\}
\thanks{This research is supported by the Australian Government Research Training Program (RTP) Scholarship. The authors are with the School of Engineering, RMIT University, Melbourne, VIC 3000, Australia }
}

\maketitle
\begin{abstract}
Through Non-Terrestrial Networks (NTN), a global coverage connecting areas with minimal or no terrestrial services is envisaged using satellite and air-borne platforms. However, one of the challenges to delivering NTN communications is the increased levels of path-loss, relative to typical terrestrial scenarios, due to the vast communication distance towards satellite platforms. Typically, cellular IoT technologies adopt frame repetitions where many redundant data transmissions are included to enhance its success rate. This can be also extended to satellite links to counter the effect of long distance communications on the link budget. In this paper, we put forth an analytic framework that captures the repetition behavior based on the probability of line-of-sight (LoS), as it heavily influences the propagation conditions. We analyze the coverage performance in terms of the frame success rate, which allows us to tune the repetitions behavior to suit a given satellite admittance region. Furthermore, we obtain an optimal global average success rate by considering the availability of satellites. The presented framework can facilitate coverage enhancements for IoT-over-Satellite networks.

\emph{Keywords - IoT-over-satellite, frame repetitions, Non Terrestrial Networks, stochastic geometry, satellite networks}
\end{abstract}

\section{Introduction}
The current era is witnessing a massive surge in IoT adoptions arising from a wide range of applications such as, smart farming, logistics, environmental monitoring, and smart cities. The number of cellular IoT subscriptions are expected to reach a staggering number of approximately 5 billion by 2026~\cite{saterric}. To support the increasingly ubiquitous IoT applications, it is required to have a network offering an extended worldwide coverage~\cite{sat2D}. So far, the traditional terrestrial network has been limited to providing robust connectivity to urban, suburban and some rural areas. However, due to the lack of economical, and sometimes technical, feasibility such terrestrial networks fell short in providing such an envisioned global coverage. As such, Third Generation Partnership Project (3GPP) included Non-Terrestrial Networks (NTN)~\cite{sat3gpp} as part of 5G road-map towards next generation hybrid networks~\cite{basselcommletter,A2}. The NTN comprises space and aerial platforms, i.e., satellites and UAVs, that act either as relaying nodes or as stand-alone base-stations, i.e., gNB, and hence paving the road for global coverage services. However, the application of IoT over NTN comes with many challenges. The NTN channel has much higher path-loss when compared to terrestrial, where most of the propagation path is under free space condition and only interacts with a clutter layer, terrestrial scatterers, just before reaching the ground user. This is especially applicable for lower frequencies suitable for some IoT applications in the L and S bands. This is also quite different than terrestrial channels where all the propagation happens within the clutter layer. As such, LoS probability plays a major role in determining the radio channel conditions. In turn, the LoS probability is highly dependent on the elevation angle~\cite{satpaper2}, which is the angle between the satellite and the horizon as seen by the ground user. The LoS dependency in an NTN channel impacts the signal power and coverage probability.\\
One of the attributes adopted by cellular IoT access technologies, specifically by Narrow-band Internet-of-Things (NB-IoT), is to improve the coverage by means of frame repetition. In such mechanisms, devices are configured to repeat a message multiple times to improve the detection probability at the receiver. As a result, devices that operate under poor coverage conditions tend to transmit more frame repetitions than those under better conditions. Multiple works evaluate the performance of repetition mechanism on terrestrial network~\cite{ruki,Narayanan}. Moreover, the evaluation of IoT over satellite access technologies has been proposed in several good works such as~\cite{satnbiotsat,conti,A1}, where work in ~\cite{satnbiotsat} analyzes the performance of NB-IoT and addresses the challenges related to delay, while the work in~\cite{conti} presents the link budget analysis for IoT devices. However, many challenges remain unresolved such as the impact of interference induced by large numbers of repetition and the performance of frame repetition in satellite radio channels. \\
In this paper, we leverage the concept of frame repetition, which is already deployed in several terrestrial IoT access technologies, and extend its application in IoT-over-satellite under realistic satellite-to-ground radio propagation conditions. We further evaluate the coverage probability of IoT devices in the admittance area of a satellite, while implementing repetition scheme. The contribution of this paper is the use of tools from stochastic geometry for analytically formulating a framework that models frame repetitions in terms of probability of line-of-sight. The paper further demonstrates the ability to tune the repetition factor and the admittance region such that an optimal repetition profile maximizes the global average success rate while considering the satellite availability. 
\section{System Model}
\subsection{Geometric Model}
\begin{figure}[t]
\centering
\includegraphics[width=0.8\linewidth]{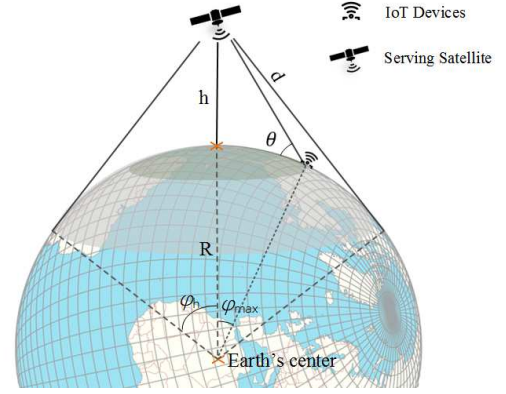}
 \caption{Illustration of a satellite admittance region, zenith angles and the IoT device at a given elevation angle.}  \label{fig:sat}
\end{figure}

Consider a satellite in a circular orbit at an altitude $h$ from the mean sea level. Ground IoT devices are assumed to be distributed by an isotropic and homogeneous Poisson Point Process (PPP) on the Earth's surface. An important parameter that impacts the performance is the angle formed between the satellite point, the Earth center, and the observer, this angle is termed as zenith angle, denoted $\varphi$.The elevation angle~$\theta$, which is the angle between satellite and horizon seen by ground user, can be obtained from the zenith angle using geometric reasoning as follows~\cite{basselcommletter}
\begin{align} \label{1}
\theta= \mathrm{acot}\left(\frac{\sin~\varphi}{\cos~\varphi-\alpha}\right), ~\text{where}~\alpha=\frac{R}{h+R}.
\end{align}
The domain of the elevation angle is $\theta\in[0,\frac{\pi}{2}]$, where $\theta=0$ represents the case when the satellite is at the horizon and $\theta=\pi/2$ is when the satellite is at the device zenith. Accordingly, $\varphi\in[\varphi_{\mathrm{h}},0]$ $\rightarrow$ $\theta\in[0,\frac{\pi}{2}]$, where $\varphi_{\mathrm{h}}$ is the maximum zenith angle formed at the horizon. The scenario is depicted in Fig.~\ref{fig:sat}. It can be shown that the maximal horizon zenith angle is given by~\cite{satpaper1}, $\varphi_{\mathrm{h}}=\text{acos}~\alpha$.
As such, the serving satellite can be geometrically seen from any point in a certain spherical cap region. However, note that IoT devices that have very low elevation angle are deemed to have an extremely poor radio link performance as a result of deep path-loss. Consequently, the probability of successful transmission in such cases is low. As such, we define a~\textit{minimum elevation angle}, denoted as $\theta_\mathrm{min}$, which limits the admittance region of IoT devices. There are multiple ways to limit the admittance region practically; for example by controlling the beamwidth of the satellite or by programmaticaly preventing IoT devices from transmitting when the satellite is below the threshold elevation angle, i.e., $\theta \leq \theta_{\mathrm{min}}$. Due to this impact, the zenith angle becomes lower than $\varphi_{\mathrm{h}}$ and thus limited by the general threshold $\varphi_{\mathrm{max}}$, i.e. $\varphi\in[0,\varphi_{\mathrm{max}}]$ $\rightarrow$ $\theta\in[\frac{\pi}{2}, \theta_{\mathrm{min}}]$. The relation between these two angles is given by \eqref{1} as follows
\begin{align}
    \theta_\mathrm{min}= \mathrm{acot}\left(\frac{\sin~\varphi_\mathrm{max}}{\cos~\varphi_\mathrm{max}-\alpha}\right), ~\text{where}~\alpha=\frac{R}{h+R}.
\end{align}
Due to the near ground objects and foliage, i.e., also called \textit{clutter}, geometric LoS is sometimes blocked by these obstacles. The probability of line-of-sight $p_{\LoS}$ is an important parameter that heavily influences the path-loss and the probability of successful transmission. Therefore, we adopt the LoS model developed in~\cite{satpaper1}. However, we re-write the $p_{\LoS}$ in terms of $\varphi$ as follows
\begin{align}
   p_{\LoS} &= \exp\left(-\beta \cot\theta\right)= \exp\left(-\beta \frac{\sin~\varphi}{\cos~\varphi-\alpha}\right),
\end{align}
where $\beta$ is a controllable modeling parameter related to the underlaying propagation environment. Note that the probability of non-line-sight is given as $p_{\NLoS} = 1-p_{\LoS}$.
\subsection{Repetition Model}
\begin{figure}
     \includegraphics[width=\linewidth]{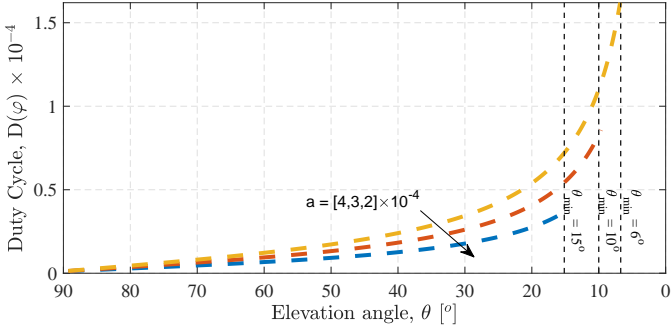}
 \caption{Effective duty cycle $D(\varphi)$ in \eqref{dc} vs $\theta$ for various combinations of tuning factor $a$ and $\theta_\mathrm{min}$.}   \label{fig: DutyCycleA}
\end{figure}

\begin{figure}
     \includegraphics[width=\linewidth]{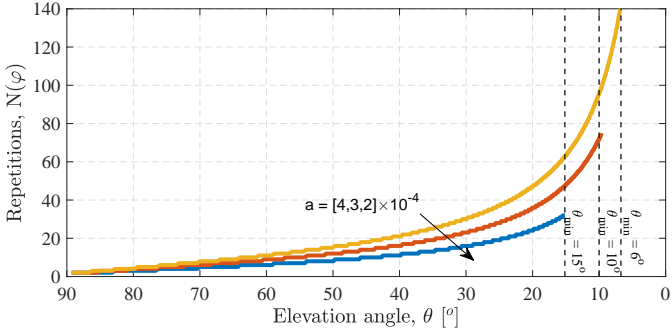}
 \caption{Repetitions $N(\varphi)$ in \eqref{rep} vs $\theta$ for various combinations of tuning factor $a$ and $\theta_\mathrm{min}$.} 
 \label{fig: DutyCycleB}
\end{figure}

Since the purpose of repetitions to improve the coverage~\cite{satnbiotism}, the devices with lower line-of-sight probability are expected to rely on this feature more often in order to achieve successful transmission. This intuitively indicates that devices with larger zenith angle, $\varphi$, will need more repetitions. Hence, the developed model in~\cite{satRepAnalysis} which primarily depends on the distance towards the terrestrial base station is adopted but with changing this dependency to the zenith angle. Assuming an initial duty cycle of $D_\mathrm{o}$\footnote{$D_\mathrm{o}$ is ratio of the time-on-air of message $T_\mathrm{m}$ to the maximum duration till next update $T_\mathrm{max}$}, the effective duty cycle, when, repetition is applied, is modeled as
\begin{equation}
  D\left(\varphi\right) = D_\mathrm{o} + (1-D_\mathrm{o})g(\varphi),
\end{equation}
where $g(\varphi)$ is a custom shaping function that satisfies the criterion $g\left(\varphi\right)\in\left[0, 1\right]$, $D\in\left[D_\mathrm{o},1\right]$, since the duty cycle cannot exceed unity. A suitable function that satisfies the above requirement is a modified version of the probability of the non-line-of-sight as follows, \begin{equation}\label{dc}
D\left(\varphi\right) = D_\mathrm{o} + (1-D_\mathrm{o})\left[1-\exp\left(-a \beta \frac{\sin~\varphi}{\cos~\varphi-\alpha}\right)\right],
\end{equation}
where $\varphi\in \left(0,\varphi_\mathrm{max}\right)$ and $a$ $\in\left[0,1\right]$ is a controllable tuning factor. Accordingly, the number of transmissions, i.e., repetitions is give by \begin{equation}\label{rep}
N\left(\varphi\right)= \left\lceil{ \frac{D\left(\varphi\right)}{D_{\mathrm{o}}}}\right\rceil.
\end{equation}
The impact of tuning factor $a$ on the duty cycle and repetitions is depicted in Fig.~\ref{fig: DutyCycleA} and Fig.~\ref{fig: DutyCycleB}, respectively. Smaller values of $a$ correspond to lower overall repetitions and vice-versa. Furthermore, the figures also illustrate the effect of the admittance angle $\theta_{\mathrm{min}}$. It is to be noted that both $a$ and $\theta_{\mathrm{min}}$ can be tuned in order to optimize the overall success probability. Due to increased transmissions as a result of repetition, the \textit{effective} device density in the admittance region becomes $\lambda(\varphi)=\lambda_{o}D\left(\varphi\right)$, where $\lambda_\mathrm{o}$ is the density of ground devices. That is, in case of no repetition, $D(\varphi) \to D_\mathrm{o}$ and $\lambda(\varphi) \to \lambda_\mathrm{o}$. A thorough explanation of repetition model is given in \cite{satRepAnalysis}.
\subsection{Zenith angle distribution}
\begin{figure}
     \includegraphics[width=\linewidth]{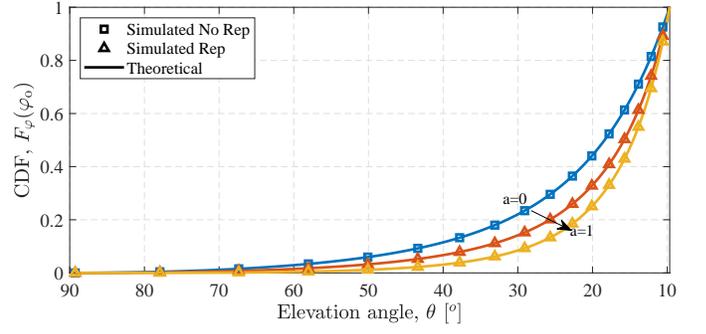}
 \caption{The CDF of the satellite zenith angle $\varphi$ in~\eqref{eqcdf} within the admittance region. Showing the results of various tuning factors $a = [ 0, 1\times10^{-5}, 1]$, at admittance $\theta_\mathrm{min}=10^{o}$.}  \label{CDF}
\end{figure}

For a given satellite admittance region defined by the angle domain, $\theta\in\left[\theta_\mathrm{min}, \frac{\pi}{2}\right)$, the users will view the serving satellite within the zenith angle domain, $\varphi\in\left(0,\varphi_{\mathrm{max}}\right)$, where $\varphi_\mathrm{max} \leq \varphi_\mathrm{h}$, as shown in Fig.~\ref{fig:sat}. As introduced earlier, the elevation angle or the equivalent zenith angle of the satellite heavily impacts the repetition rate. Thus, the analysis of satellite's zenith angle distribution is essential to obtain the probability of successful transmission. Assuming PPP, the ground devices or IoT devices are randomly placed within the admittance region. Accordingly, the probability of finding devices within a given angle $\varphi_\mathrm{o}$ is formulated as follows
\begin{align}\label{cdf}
\scriptstyle F_\varphi(\varphi_\mathrm{o})=\mathbb{P}\left(\varphi\leq \varphi_\mathrm{o}\right)
    &= \frac{\text{avg. no. of active devices in $\varphi_\mathrm{o}$} }{\text{avg. no. of active devices in $\varphi_\mathrm{max}$}}~,
\end{align}
which represents the cumulative distribution function (CDF) of the satellite zenith angle. The average number of points can be found by invoking Campbell theorem~\cite{satbook1} for a PPP. Accordingly, the average number of points in a region bounded by an angle $\left[0, \varphi\right]$ is given by
\begin{equation}\label{K}
    K(\varphi)=\int_0^\varphi 2 \pi R^2 \lambda(\varphi) \sin(\varphi)~ \mathrm{d}\varphi,
\end{equation}
because the spherical strip area is given by, $2 \pi R^2 \sin(\varphi) \mathrm{d}\varphi$. Substituting in \eqref{K} and replacing $\lambda(\varphi)=\lambda_\mathrm{o} D(\varphi)$ we obtain
\begin{equation}\label{eqcdf}
    F_{\varphi}(\varphi_\mathrm{o})=\frac{K(\varphi_\mathrm{o})}{K(\varphi_\mathrm{max})}=\frac{\int_\mathrm{0}^{\varphi_o}\sin(\varphi_\mathrm{o})  D(\varphi_\mathrm{o})~\text{d}\varphi}{\int_{0}^{\varphi_\text{max}} \sin(\varphi) D(\varphi)~\text{d}\varphi}.
\end{equation}
The probability density function (PDF) is thus obtained by differentiating \eqref{eqcdf} as follows
\begin{align}
    f_\varphi(\varphi_\mathrm{o})= \frac{\sin(\varphi_\mathrm{o}) D(\varphi_\mathrm{o})}{\int_{0}^{\varphi_\text{max}} \sin(\varphi)
    D(\varphi)~\text{d}\varphi }.
\end{align}
An illustration of the CDF at $\theta_{\mathrm{min}}=10^{\mathrm{o}}$ is shown in Fig.~\ref{CDF} along with Monte-carlo simulations to verify the analytical results. Note that the distribution changes for different tuning factor $a$.
\section{Link Performance}
Majority of IoT technologies rely on simple random access techniques in order to reduce the complexity of the IoT devices. Examples can be found in LoRaWAN and in the initial random access stage of NB-IoT. Accordingly, collision of frames is bound to happen due to the lack of coordination. 
\subsection{Channel Model}
For a given device with transmitting power $p_{\mathrm{t}}$, at an angle $\varphi_{\mathrm{o}}$, the received signal power in the uplink can be modeled as $S= p_\mathrm{t} l(\varphi_{\mathrm{o}}) \zeta(\varphi_{\mathrm{o}}) $\cite{satpaper4}, where $l(\varphi_{\mathrm{o}})$ is the reciprocal of the free space path-loss (FSPL) $\left(\frac{4 \pi d f}{c}\right)^{2}$, while $\zeta(\varphi_{\mathrm{o}})$ is the excess gain caused by ground clutter, and its distribution follows simple Gaussian mixture, this path-gain heavily depends on the probability of LoS $p_{\LoS}$ as given in eq.~(7) in~\cite{introAKS1}. 
\subsection{Interference}
\begin{figure}[t]
     \includegraphics[width=\linewidth]{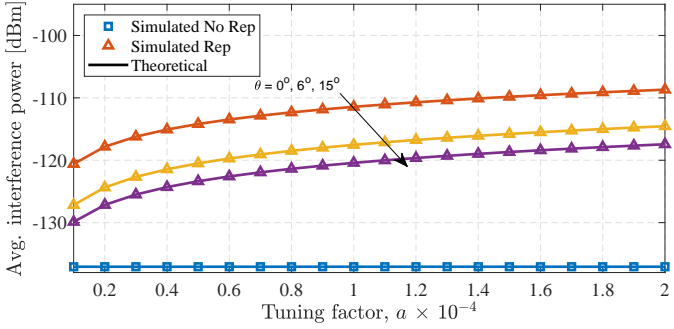}
 \caption{The average interference power given in~\eqref{In} versus the tuning factor $a$ with different admittance angles $\theta_\mathrm{min}$.}   \label{Int}
\end{figure}

For a given device located at $x_\mathrm{o}$, the simultaneous transmissions imposed by other devices within the admittance region cause interference with a total aggregated power, denoted $I$, obtained as follows
\begin{align}
    I=\sum_{x_{\mathrm{i}}\in\Phi \text{\textbackslash} x_{\mathrm{o}}} \kappa p_\mathrm{t} \zeta(\varphi_{i}) l(\varphi_{i}),
\end{align}
where $\Phi$ is the set of all devices in the admittance region, $\kappa$ is a system parameter that captures the impact of interference mitigation caused by resource coordination in the access system. Note that $\kappa=1$ indicates the worst case scenario when no coordination is implemented. Accordingly, the spatial average of the interference is obtained by invoking Campbell’s theorem of sums and is given as
\begin{align} \label{In}
    \mathbb{E}[I]=2\pi R^2 \lambda_\mathrm{o}\int_{o}^{\varphi_\mathrm{max}}\bar{\zeta}(\varphi)l(\varphi)\sin(\varphi) D(\varphi) \mathrm{d}\varphi
\end{align}
where $\bar{\zeta}$ is the average excess path-gain expressed as~\cite{satpaper2}
{\small
\begin{align}
   \bar{\zeta}= p_{\LoS}\exp\left(-\mu_{\LoS}+\frac{\sigma^2_{\LoS}}{2}\right) + p_{\NLoS}\exp\left(-\mu_{\NLoS}+\frac{\sigma^2_{\NLoS}}{2}\right),
\end{align}}
where $\mu$ and $\sigma$ are the mean and standard deviation parameters associated with respective LoS and non-LoS components. To validate the above analytic results, we depict in Fig.~\ref{Int} a comparison between the analytic results of \eqref{In} and Monte Carlo simulations at various values of $a$ and $\theta_{\mathrm{min}}$. Note the increase in the aggregated interference power when deploying the repetition scheme. Nevertheless, it is shown in the next subsection that the probability of success can be further increased with careful tuning of $a$ and $\theta$.
\subsection{Probability of success}
\begin{figure}
\includegraphics[width=\linewidth]{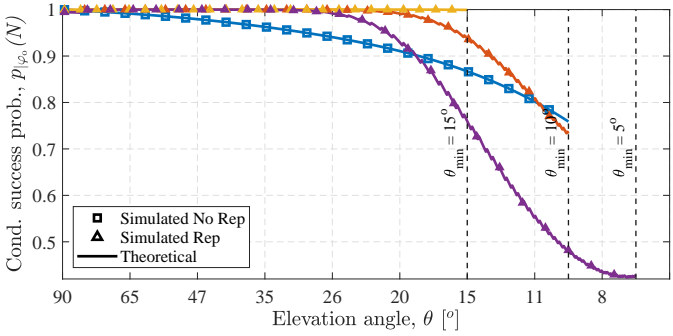}
\caption{Conditional success probability in~\eqref{sinr1} at $a=2\times10^{-4}$, $\gamma=-10$, and $\theta_{min}=\left[5^\circ, 10^\circ, 15^\circ\right]$ for repetition scenario, and at $\theta_{min}=\left[10^\circ\right]$ for no repetition scenario.}  \label{avgsuccess}
\end{figure}

\begin{figure}
     \includegraphics[width=\linewidth]{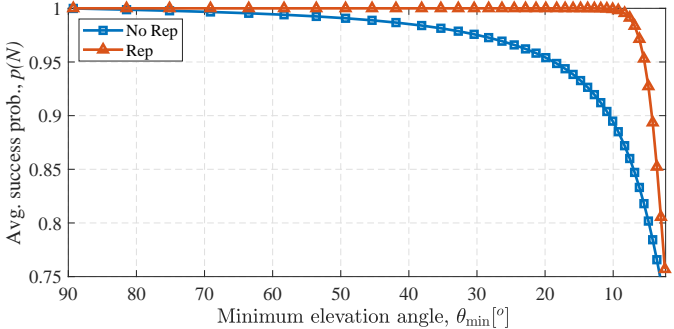}
 \caption{The average success probability in~\eqref{pc} within a single satellite admittance region, at $a=5\times10^{-5}$ and $\gamma=-10$}\label{figpc}
\end{figure}

To measure the probability of success under influence of frame repetition, we utilize a vital metric of communication link which is the ratio of received signal power to the interference plus noise power (SINR). The probability of a frame having an SINR greater than a given design threshold $\gamma$, is equivalent to the frame success probability. For a given IoT device viewing the satellite at a zenith angle $\varphi_\mathrm{o}$, the success probability of a single transmission is formulated as follows
{\small
 \begin{align}
  p_{|\varphi_\mathrm{o}}(1)&=\mathbb{P}\left(SINR>\gamma\right)=\mathbb{P}\left(\frac{S}{\bar{I}+W}>\gamma\right)\nonumber\\
  &=\mathbb{P}\left({\zeta}>\frac{\gamma(\bar{{I}}+{W})}{\mathrm{p}_\mathrm{t}l(\varphi)}\right)=1-F_\mathrm{\zeta}\left(\frac{\gamma(\bar{{I}}+{W})}{\mathrm{p}_\mathrm{t}l(\varphi)}\right),
\end{align}}
where ${W}$ is the average thermal noise power, $\bar{I}$ is given in~\eqref{In} and $F_\mathrm{\zeta}$ is given as~\cite{basselcommletter}
\begin{align}
\scriptstyle F_\mathrm{\zeta}= p_{\LoS}F_\mathrm{X}\left(x,-\rho\mu_{\LoS}, \frac{\sigma_{\LoS}^2\rho^2}{2}\right)+p_{\NLoS}F_\mathrm{X}\left(x,-\rho\mu_{\NLoS}, \frac{\sigma_{\LoS}^2\rho^2}{2}\right),
\end{align}
where $F_\mathrm{X}$ is the CDF of the log-normal distribution as ${F_\mathrm{X}(x)=\frac{1}{2}+\frac{1}{2}\text{erf}\left(\frac{\ln x-\mu}{\sigma \sqrt{2}}\right)}$ and $\rho=\frac{\ln 10}{10}$. For a device attempting its transmission $N$ number of times, the probability of at least one of the frames to be successfully received is formulated as
\begin{align} \label{sinr1}
   p_{|\varphi_\mathrm{o}}(N)=1-[1-p_{|\varphi_\mathrm{o}}(1)]^{N}.
\end{align}
We illustrate the success probability in Fig.~\ref{avgsuccess} with respect to a range of elevation angles for three given examples of the admittance region. We note that for smaller admittance region, i.e., larger values of $\theta_{\mathrm{min}}$, the probability of success increases. For small $\theta_{\mathrm{min}}$ the increase in admittance region causes devices to incur more repetition, hence, increasing the interference level, which overshadows the performance enhancement as shown for $\theta_\mathrm{min}=5^{\circ}$. Subsequently, to find the average success probability for a given admittance region bounded by $\varphi_{\mathrm{max}}$, we decondition~\eqref{sinr1} over the distribution of $\varphi$ as follows
\begin{align}\label{pc}
    p(N)=\int_{0}^{\varphi_\mathrm{max}}p_{|\varphi_\mathrm{o}}(N) f_\varphi(\varphi_\mathrm{o})~\mathrm{d}\varphi.
\end{align}
It is illustrated in Fig.~\ref{figpc} where the success probability of the repetition scenario clearly outperforms the no repetition scenario. Note that this performance is for a single satellite spot, i.e. single admittance region. The following section extends results to a constellation of satellites. 
\subsection{Satellite Constellation Performance}
\begin{figure}
     \includegraphics[width=\linewidth]{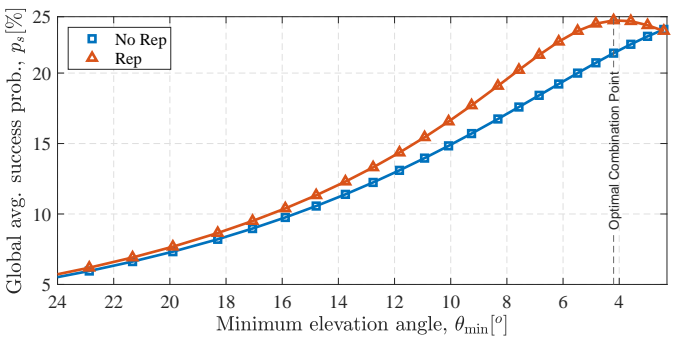}
 \caption{Global average success probability in~\eqref{ps} evaluated at $a=5\times10^{-5}$ and $\gamma=-10$, $k=10$.}\label{optztheta}
\end{figure}

Assuming most of the satellites are positioned in a non-GEO (Geosynchronous Equatorial Orbit), which are in a relative motion with respect to the ground IoT devices. As such, when reducing the size of the admittance region, the probability of having the coverage spot of the device is reduced. Accordingly there is a trade-off between satellite availability and its success rate within the spot. To capture this trade-off, we extend the performance by including the satellite availability, which is the ratio of satellite admittance area to the total surface area of Earth's sphere. For a small number of satellites, denoted $k$, and assuming non-overlapping spots, then the probability that a device is within the coverage is given as
\begin{align}
p_{\mathrm{spot}}
&=\frac{k~2\pi R^2\left(1-\cos\varphi_{\mathrm{max}}\right)}{4\pi R^2}=\frac{k\left( 1-\cos\varphi_{\mathrm{max}}\right)}{2}, 
\end{align}
where $0\leq p_{\mathrm{spot}} \leq 1$. Accordingly, the probability of a successful frame delivery in a constellation of $k$ satellite with non-overlapping spots is given by
\begin{align} \label{ps}
  p_{\mathrm{s}} = p_{\mathrm{spot}} \times p(N).
\end{align}
We depict in Fig.~\ref{optztheta} the global average success probability evaluated for one tuning factor and various minimum elevation angles. We can clearly note the trade-off induced by adjusting the admittance region, i.e. by controlling $\theta_\mathrm{min}$. Moreover, as the shape of the curve is impacted by the tuning factor $a$, an optimal point for a given admittance angle exists. In the following section we shed light on the joint optimization of these two parameters.
\section{Discussion on Results}
\begin{figure}[ht]
\includegraphics[width=\linewidth]{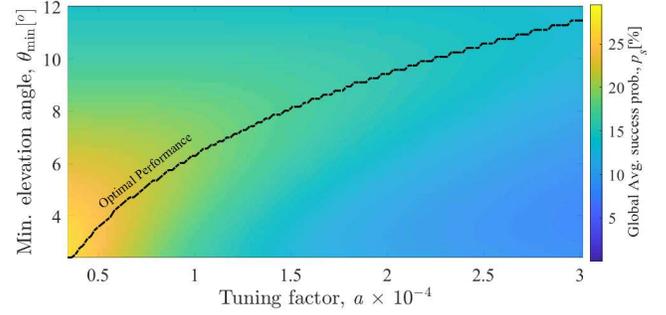}
\caption{Global average success probability in~\eqref{ps} calculated for various $\theta_{\mathrm{min}}$, $a$ and at $k=10$.}  \label{optz3}
\end{figure}

In order to understate the joint impact of the admittance region size and the tuning factor on the global coverage probability, we depict in Fig.~\ref{optz3} the global success probability while varying factor $a$ and admittance angle $\theta_{\mathrm{min}}$. The solid line represents the optimal combination values for both parameters. Other factors that impact this optimal solution are related to the initial density of the ground devices $\lambda_\mathrm{o}$ and the propagation conditions manifested in channel parameters since these parameters influence \eqref{ps}. As such, network designer can optimize the admittance region and tuning factor based on the operating parameters. To validate our analytical results we utilize Monte Carlo simulations that consider non-overlapping circular orbit satellites. Other simulation parameters include a satellite altitude of 550~km, noise average power of -138~dBm~\cite{satpaper2}, center frequency at 2000~MHz, device $p_\mathrm{t}$ at 23~dBm as utilized in many low power IoT devices, and a bit rate of 8~Kbps resulting in $D_{\mathrm{o}}$ of $1\times10^{-6}$. 
\section{Conclusion}
This paper presented an analytic model that captures the performance of IoT-over-satellite in the uplink for non-GEO orbits using NTN radio channel. The model employs frame repetitions to enhance the success rate of transmissions. A trade-off between the satellite coverage region, admittance region, and satellite availability is formulated in order to evaluate the performance of a constellation of satellites. Accordingly, we found an optimal combination of parameters that maximizes the frame success rate. These results can facilitate network design and optimization for IoT-over-satellite links.
\bibliographystyle{IEEEtran}
\bibliography{Manzoor_WCL2021_1889.bib}
\end{document}